\documentclass[12pt]{article}

\usepackage{amsmath}
\usepackage{amsfonts}
\usepackage{amssymb}
\usepackage[english,russian]{babel}

\newcommand{\be}{\begin{equation}}
\newcommand{\ee}{\end{equation}}

\title{Gravitational interaction for light-like motion in classical
and quantum theory}
\author{Nikolai V. Mitskievich}
\date{~}
\begin{document}

\selectlanguage{english}

\maketitle
\begin{abstract}
On the basis of an exact vacuum solution of Einstein's equations,
{\it vis}. the pencil-of-light field, we study the light-like
motion of test and non-test objects. We also consider the quantum
theoretical interaction of massless scalar particles through
virtual gravitons. The dragging phenomenon is manifested and its
agreement with astronomical observations established.\\
This paper submitted to {\bf arXiv} is a somewhat reedited copy of
my article dedicated to Dr. Ivar Piir in a volume published on the
occasion of his 60th birthday in 1989 in Tartu by the Estonian
Academy of Sciences.
\end{abstract}
\section{Introduction}
The interplay of classical and quantum physics is very profound.
On the one hand, the classical Poisson brackets are often
heuristically replaced by commutators, or similar procedures are
performed. On the other hand, the Ehrenfest theorem helps to
``deduce'' the classical theory from the quantum one; there also
exist other forms of the correspondence principle. One of these
was noted by I.Piir \cite{1}: it related the cross-section of
scattering of a photon on the Schwarzschild centre with the
classical light bending effect in the Schwarzschild field. This
analysis, published as early as in 1957, retains its fundamental
importance even today. Continuing the study, I extended these
results to include this scattering of other particles, as well as
to make use of only ``one half'' of the correspondence principle
limit, {\it i.e.} when one of the two mutually scattered particles
became a classical centre, while the other continues to be a
quantum theoretical object (see \cite{2,3}).

Another trend of research in classical General Relativity at all
times was to study dragging effects, beginning with Einstein's
admission of Mach's principle \cite{4}, and the approximate
results of Thirring and Lense \cite{5,6}. Usually, the dragging
effects are related to the rotation of the gravitational field
source (this reflects the idea of Mach's principle); however,
similar and not less striking phenomena are also predicted for
motion in the NUT field \cite{7} and in the pencil-of-light field
\cite{8}. Neither of these fields is associated with the rotation,
but they, and also the Kerr field, have exact electromagnetic
analogues, {\it vis.} the characteristic magnetic or
quasi-magnetic fields. Therefore one could treat the dragging
phenomenon as a totality of quasi-magnetic or -electric effects in
the gravitation theory.

However, the quantum theoretical manifestations of dragging are
practically unknown (not mentioning the spin-orbital type effects
for the general relativistic Dirac field, namely effects in the
primarily quantized theory; see \cite{9,3}). I dare to try to fill
this gap and give here new examples of the classical and quantum
theoretical parallels in this field, taking the opportunity of the
60th birthday of my dear friend and highly respected colleague
Ivar Piir, in the hope that he would enjoy reading here some lines
on the subjects of quantum and classical theory of gravitation,
the theory so close and exciting for both of us.

An approximate expression of the gravitational field of a
light-like linearly extended object was found in 1931 (my
birth-year) by Tolman, Ehrenfest, and Podolsky \cite{10}. An exact
solution for this field of a pencil of light \cite{8} turned out
to be a special case of the well-known Peres' wave \cite{11}. This
is in fact not a wave, the field is atationary, though it belongs
to the Petrov type N. In our discussion of the classical dragging
phenomena we shall consider a further generalization of this field
to a case when the ``pencil'' possesses angular momentum oriented
along it (which is the spatial direction of its light-like motion)
\cite{12}: \be \label{1} ds^2=2dv(du+\kappa\ln\sigma\rho
dv+gd\varphi)-\rho^2d\varphi^2-d\rho^2. \ee Here $\kappa=8\gamma
\epsilon$, $\gamma$ being the Newtonian gravitational constant and
$\epsilon$, the linear energy density along the $z$ axis; the
function $g(v)$ ($\frac{dg}{dv}\neq 0$) describes the angular
momentum of the source. In this vacuum solution $\kappa$ may
depend on $v$, but here we assume it to be constant for the sake
of simplicity (but for constant $g$ the angular momentum
vanishes).

\section{A classical case of the light-like motion}

Now we take $ds^2=0$ and denote the affine parameter along
geodesics by $\lambda$. The motion of a test particle is described
by the geodesic equations, \be \label{2} \frac{d}{d\lambda}\left(
g_{\alpha \beta}\frac{dx^\beta}{d\lambda}\right)=\frac{1}{2}
g_{\mu\nu,\alpha} \frac{dx^\mu}{d\lambda}\frac{dx^\nu}{d\lambda},
\ee which are readily integrated for $\alpha$ labelling the
variables $v$ and $\varphi$: \be \label{3} v=A\lambda+B, ~ A={\rm
const}, ~ B={\rm const} \ee (the second integral of motion) and
\be \label{4} \frac{d\varphi}{d\lambda}=\frac{Ag+C}{\rho^2}, \ee
where $C=$const (the first integral). The equations which
correspond to the variables $u$ and $\rho$, are \be \label{5}
\frac{d^2u}{d\lambda^2}=-g\frac{d^2\varphi}{d\lambda^2}-\frac{2
\kappa A}{\rho}\frac{d\rho}{d\lambda} \ee and \be \label{6}
\frac{d^2 \rho}{d\lambda^2}=-\frac{\kappa A^2}{\rho}+\frac{(Ag+
C)^2}{\rho^3}. \ee $ds^2=0$ is itself a first integral of motion;
it reads \be \label{7} 2A\frac{du}{d\lambda}=-2\kappa A^2\ln\sigma
\rho-\frac{A^2g^2}{\rho^2}+\frac{C^2}{\rho^2}+\left(\frac{d\rho}{d
\lambda}\right)^2. \ee One may replace by it one of the above
equations, or use it in order to check the results on which the
equation (\ref{7}) imposes a certain constraint (the light-like
motion case).

We consider first the simplest cases:\\
$1^{\rm o}$. Let $\varphi=$const.; then equation (\ref{4}) yields
$A=0=C$. The equation (\ref{7}) itself reduces to $\frac{d\rho}{d
\lambda}=0$, {\it i.e.} $\rho=$const., and from (\ref{5}) we get
$u=K\lambda+L$, $K$ \& $L$ being constants.\\
$2^{\rm o}$. Let $\rho=$const. Then from (\ref{6}) it follows that
$\frac{d\varphi}{d\lambda}=\frac{\sqrt{\kappa}A}{\rho}=$const.,
which can be realized only if $A=0$ (since we admit $\frac{dg}{dv}
\neq 0$).The problem is thus reduced to the first case.

This shows absolutely no influence of the spinning pencil of light
upon a test photon moving parallel (not antiparallel!) to it. The
picture becomes more obvious if we put $\sqrt{2}u=t+z$ and
$\sqrt{2}v=t-z$ ({\it cf.} \cite{8,12}).

The other cases describe photons moving non-parallel or
(instantaneously) antiparallel to the pencil of light. The most
interesting case among these would be when $\frac{d\rho}{d\lambda}
$ and $\frac{d\varphi}{d\lambda}$ both vanish at (say) $\lambda=
0$, $A\neq 0$:\\
$3^{\rm o}$. Let us choose fairly arbitrarily $\rho=f(\lambda)$
with $\left.\left(\frac{df}{d\lambda}\right)\right|_{\lambda=0}$.
Now a substitution of $\rho$ into equation (\ref{6}) determines
the function $g(v)$ (with the help of equation (\ref{3})). The
equation (\ref{4}) then yields $\phi(\lambda)$ {\it via} a
straightforward integration, and the equation (\ref{5}) yields
$u(\lambda)$ as well. Constraints on the constants which enter the
integrals emerge after substitution of them into (\ref{7}). A
further adjustment of the function $f(\lambda)$ should now be made
in order to meet the condition with $\left.\left(\frac{df}{d
\lambda} \right)\right|_{\lambda=0}$. As an example of an adequate
choice of the function $f$ we can produce here:
$$
\rho=\rho_0\left[1+\frac{\kappa}{2}\left(\frac{A\lambda}{\rho_0}
\right)^2\right]^{-1}.
$$
Then
$$
\frac{d\varphi}{d\lambda}=\lambda\rho\kappa A^2C^3\left[1+
\frac{\kappa}{4}\left(\frac{A\lambda}{\rho_0}\right)^2\right]^{1/2}
\times\left[2+\kappa\left(\frac{A\lambda}{\rho_0}\right)^2+
\frac{\kappa^2}{4}\left(\frac{A\lambda}{\rho_0}\right)^4
\right]^{1/2}.
$$
This solution describes both the gravitational attraction and
angular dragging in the field of a spinning pencil of light for
the case of non-parallel motion of a test photon. No interaction
in the case of parallel motion ($1^{\rm o}$ \& $2^{\rm o}$) gives
a manifestation of dragging too, though a peculiar one; this case
seems to be very important since it corresponds to exact
non-interaction in a self-consistent problem (see \cite{8,12}).
This means that a superposition of fields of two parallel pencils
of light is also an exact vacuum Einstein field, and it generally
admits only one (null) Killing vector, $\partial_u$.

\section{Light-like motion in quantum theory}

Within the scope of the quantum theory, we shall consider here
interaction of two massless particles through virtual gravitons.
Examples of such calculations can be found in \cite{3}, Section
7.3, though one has to set the rest masses equal to zero in the
respective formulae, so that I would have no need to consider in
\cite{3} the case of a parallel motion of incoming particles. If
only two particles are present, we may transform the picture ({\it
via} a Lorentz transformation) so that the particles will move
either parallel or antiparallel to each other. The case of the
parallel motion is a very special one, since it remains invariant
under all transformations, while antiparallel, on the contrary, is
a more general one. The latter case admits a further
transformation to the centre-of-mass frame which was considered in
\cite{3}. However, the laboratory frame of Ref. 3 becomes now
futile, since one of the particles there has to be supposed to be
at rest which cannot hold when its rest-mass vanishes. Hence the
parallel motion should be considered in a general frame, and the
antiparallel motion too, because we are interested in a final
transition to the scattering by a classical field.

We suppose the interacting particles to be the scalar ones, but
the scalar fields to be distinguishable; this will simplify the
calculations. The standard two-vertex diagram of Fig. 7 in
\cite{3} describes the process under consideration where $r$ and
$s$ are the four-momenta of the incoming particles, and $p$ and
$q$ those of the outgoing ones.

We shall use the expression (7.3.4) for the differential
cross-section \be \label{8} d\sigma=(2\pi)^2{p_0}^2|F|^2\left[
\left|\frac{\vec{r}}{r_0}-\frac{\vec{s}}{s_0}\right|\times\left|1+
\frac{\partial q_0}{\partial p_0}\right|\right]^{-1}, \ee where
the matrix element is (7.3.3), \be \label{9} F=\frac{i\kappa^2[
(p\cdot r)(q\cdot s)-(p\cdot q)(r\cdot r)]}{4(2\pi)^2(p_0q_0r_0s_0
)^{1/2}(p-r)^2}, \ee $\kappa=\sqrt{2\varkappa}$, $\varkappa$ being
Einstein's gravitational constant, and the conservation law
$p_nu+q_\nu=r_\nu+s_\nu$ is to be taken into account, as well as
relations like ${p_0}^2=\vec{p}^{\,2}$. It is easy to check that
$p\cdot r=q\cdot s$, $p\cdot s=q\cdot r$, $p \cdot q=r\cdot s$.
Then
$$
(p\cdot r)(q\cdot s)-(p\cdot q)(r\cdot s)-(p\cdot s)(q\cdot r)=
-2(p\cdot s)(p\cdot q).
$$

In the case of antiparallel motion the conservation law gives
${q_0}^2=\vec{q}^{\,2}=(r_0-s_0)^2-{p_0}^2+2p_0q_0$, so that
$\frac{\partial q_0}{\partial p_0}=1$, since $\frac{\partial r_0}{
\partial p_0}=\frac{\partial s_0}{\partial p_0}=0$ by definition.
Moreover, $|\vec{r}/r_0-\vec{s}/s_0|=2$. If we denote $\vec{r}
\cdot \vec{p}/r_0p_0=\cos\theta$, then $r\cdot p=2r_0p_0\sin^2(
\theta/2)$, $s\cdot p=2s_0p_0\cos^2(\theta/2)$ and $F=i\kappa^2
s_0\left[2( 2\pi)^2r_0\right]^{-1}\cot^2(\theta/2)$. Hence, \be
\label{10} d\sigma=\frac{\kappa^4{p_0}^2s_0}{(8\pi)^2r_0}\cot^4
(\theta/2)d\Omega, \ee which gives in the center-of-mass frame
exactly the same expression which the equation (7.3.12) will yield
if $m=M=0$:
$$
d\sigma_{\rm C}=\kappa^4{p_0}^2(8\pi)^{-2}\cot^4(\theta/2)d\Omega.
$$

On the other hand, in the case of parallel motion both the
numerator and denominator in (\ref{9}), as well as $|\vec{r}/r_0-
\vec{s}/s_0|$, vanish. One can cure this indeterminacy by starting
with non-zero rest masses and passing to the limit of zero ones.
Then $d\sigma_p=0$.

One can now pass to scattering on a classical point-like particle
moving with the velocity of light by taking $q_0=s_0\gg p_0=r_0$.
The cross-sections then become \be \label{11} d
\sigma_\rightleftarrows=\frac{ \kappa^4{s_0}^2}{(8\pi)^2}\cot^4(
\theta/2)d\Omega, ~ d\sigma_\rightrightarrows=0, \ee which mean
that the interaction between two light-like objects (with zero
rest-masses) in a head-on collision ($\rightleftarrows$) is four
times greater than between a beam of light and a Schwarzschild
``point'' mass at rest (see \cite{1,2,3}), \be \label{12} d
\sigma_{\rm Schw}=\frac{\kappa^4 m^2}{(16\pi)^2}\cot^4(\theta/2)
d\Omega, \ee whereas two classical objects, two quantum
theoretical ones, or a pair consisting of one classical and one
quantum object, moving in one and the same direction
($\rightrightarrows$) with the velocity of light (thus having no
rest-masses), do not interact at all. If both objects are
non-relativistic, we have to make use of the equation (7.3.14) in
\cite{3} with $p_0=m$; the corresponding cross-section becomes
another four times smaller than that of (\ref{12}) (there enters
also a difference in the angular dependence, but it is
insignificant for a scattering in small angles).

\section{Concluding remarks}

In our simple calculations above we have shown the existence of a
far-reaching harmony between quantum and classical physics also in
the ultra-relativistic region, where a peculiar property of
absolute non-interaction between any number of light-like
particles moving in parallels, is revealed. This property is very
natural from the standpoint of relativistic causality, and it
leads to a remarkable fact of exact superimposition ability of
corresponding gravitational fields in general relativity. See also
a different approach developed by Bonnor \cite{14}.

There exists even a very sensitive test which indisputably proves
the theoretical prediction of non-interaction between light-like
objects moving in parallels. This is the astronomical observation
of optical images of distant galaxies, which appears completely
undisturbed, however far the emitters of radiation were located
(except of the influence of the interstellar non-relativistic
matter).

It is necessary to stress again (see \cite{3}, p. 249) the
invalidity of the assertion that the limit of a classical centre
is achieved when the field intensity tends to infinity (see
\cite{13}, p. 191 in the Russian edition $\Rightarrow$ pp.
260---262 in the English edition); in reality, it is the property
of Newtonian inertia {\bf(energies and/or rest masses)} of quanta
of the corresponding field {\bf(not its intensity)}, that should
grow `infinitely' in order that it would be possible to pass to
scattering of particles on the corresponding classical ``centre''.

\section*{Acknowledgement}

I am grateful to Dr. Ivar Piir for the inspirations he has always
incited in me and which have always led to fascinating studies of
the enigmas of Nature. Palju, palju \~{o}nne, edu ja
kordaminekuid!

\end{document}